\documentclass{article}
\usepackage{fleqn,espcrc2}

\usepackage{graphicx}
\usepackage{url}
\usepackage{paralist}

\title{How to SYN in seven easy steps}

\author{T. A. Nieminen and J. E. R. Ross
	\address{Department of Physics, The University of Queensland
	QLD 4072, Australia}
}

\begin{document}

\begin{abstract}
The calculation of expected spectral line strengths and profiles is a
powerful tool for the analysis of the solar atmosphere, and other
stellar atmospheres. We present here a recipe in seven easy steps for
the development of such spectral synthesis software.

\vspace{-6.5cm}
{\small\noindent
\textbf{Preprint of:}\\
T. A. Nieminen and J. E. R. Ross\\
``How to SYN in seven easy steps''\\
\textit{Computer Physics Communications} \textbf{142}, 160--163 (2001)
}
\vspace{5.2cm}


\vspace{1pc}
\end{abstract}

% typeset front matter (including abstract)
\maketitle

\section{Introduction}

The solar spectrum can be calculated by (numerically)
solving the radiative transfer equation:
\begin{equation}
\mu \frac{dI_\lambda}{d\tau_\lambda} = I_\lambda - S_\lambda
\label{transfer}
\end{equation}
If we know the source function $S_\lambda$, then finding the emergent
spectrum is straightforward. With a model solar atmosphere, we have the
LTE source function, $S_\lambda(z) = B_\lambda(z)$, the Planck
radiation function, as a function of physical depth.
Unfortunately, it is not sufficient to know $S_\lambda$
as a function of physical depth $z$; we must convert the physical depth
scale to an optical depth scale:
\begin{equation}
\tau_\lambda(z) = \int_z^\infty \rho(z') \kappa_\lambda(z') dz'
\end{equation}

This requires calculation of the opacity $\kappa_\lambda$; this is the
core task of a spectral synthesis program. The opacity can be considered
to consist of a continuous opacity, and opacity due to atomic line
transitions. The main contribution to the continuous opacity is due to
bound-free and free-free H$^-$ transitions; other major sources (H$_2^+$
molecule, photoionisation of metals) also need to be considered. The
line opacity can be found from the line strength and line profile for
each line in the spectral region of interest.

\begin{figure}[htb]
\rule{7.5cm}{0.5mm}
\begin{compactenum}
\item Calculate $S_\lambda(z)$
\item Calculate continuous opacity
\item Calculate line opacities.\\
	This requires:
	\begin{compactitem}
        \item line strengths
        \item damping
        \item thermal and mass motion
	\end{compactitem}
\item Calculate optical depth scale
\item Effect of intermediate scale mass motion
\item Numerical integration of radiative transfer equation
	(\ref{transfer})
\item Non-plane parallel corrections---large scale mass motion
\end{compactenum}
\rule{7.5cm}{0.5mm}

\vspace{-8mm}
\caption{Recipe for spectral synthesis}
\label{recipe}
\end{figure}

\section{Model atmospheres}

The solar atmosphere is highly stratified; its physical
properties vary strongly with height within the
photosphere. Thus, the photosphere can be well
represented by a plane-parallel \emph{model atmosphere}. Many model
atmospheres are available, with those by Holweger and
M\"{u}ller~\cite{holmul} and Kurucz~\cite{kurucz} being commonly used.

A model atmosphere tabulates temperature, pressure, electron pressure,
density, and opacity at 5000~\AA .

With this data, and assuming LTE conditions, then the source function is
simply equal to the Planck radiation function:
\begin{eqnarray}
S_\lambda (z) & = & B_\lambda (T(z))
\nonumber \\
 & = & \frac{2hc^2}{\lambda^5} (\exp(-hc/\lambda kT) - 1 )^{-1}
\end{eqnarray}

\section{The continuous opacity}

The continuous opacity (maximum at 8200~\AA ) is dominated by the H$^-$
ion, due to its large interaction cross-section. We can also consider
other abundant absorbers. If we are only considering a small spectral
region, we can assume that the continuous opacity is uniform. In
general, we use approximation formulae and table lookup~\cite{thesis}.

The H$^-$ opacity is due to both bound-free
($\mathrm{H}^- + \mathrm{photon} \leftrightarrow \mathrm{H} + e^-$)
and free-free
($\mathrm{H} + e^- + \mathrm{photon} \leftrightarrow \mathrm{H} + e^-$)
processes. Approximation formulae for both processes are given by
Gingerich~\cite{HSCSA}

Due to its high abundance, we should also
consider bound-free and free-free absorption due to neutral hydrogen,
which depend on the Gaunt factors $g_{\mathrm{II}}$
and $g_{\mathrm{III}}$. Approximation formulae for $g_{\mathrm{II}}$ and
$g_{\mathrm{III}}$ are given by Mihalas~\cite{mihalas}.

Other sources of opacity which can be considered include
the H$_2^+$ molecule, for which cross-sections have been calculated; a
convenient tabulation for interpolation is available in~\cite{HSCSA}.
The abundances of heavier elements much lower than that of hydrogen,
and they contribute far less to the opacity. Only the
most abundant species need be considered, Mg and Si being the most
important. Calculations by Peach~\cite{peach} can be used for
interpolation.

\section{Line opacities}

The opacity due to a spectral line depends strongly on the
wavelength---which is why it's
a line---which makes it necessary to calculate the
line opacity on a sufficiently dense grid of wavelength points. For a
single (not too strong) line, a total wavelength range of $\approx$
1~\AA\ is enough; a few hundred wavelength points will be sufficient.
The use of a non-uniform grid of wavelength points can greatly speed up
line opacity calculation, at the cost of a greater effort to implement.

The opacity of a single line depends on
\begin{enumerate}
\item population of the absorber,
\item intrinsic strength of the transition,
\item and the line profile.
\end{enumerate}

The populations of absorbers depend on the atomic abundances and the
equilibrium between the atomic species, the rest of the atmosphere, and
the radiation field. We can simplify this greatly by assuming LTE.

The strength of the transition---the oscillator strength, or $f$-value
can be obtained from a suitable source of data such as the
NIST Atomic Spectra Database at
\url{http://physics.nist.gov/cgi-bin/AtData/main_asd} or the
Vienna Atomic Line Database (VALD) at
\url{http://www.astro.univie.ac.at/~vald/}.

\subsection{The line profile}

The line profile function of a stationary atom is the Lorentz profile:
\begin{equation}
\phi(\lambda) = \frac{\Delta\lambda/2\pi}
{(\lambda - \lambda_0)^2 + (\Delta\lambda/2)^2}
\label{lorentz}
\end{equation}
Thermal motions will produce a profile that is the convolution of the
Lorentz profile and the Maxwellian Doppler shift profile, giving the
Voigt profile:
\begin{eqnarray}
\phi(\lambda) & = & \frac{a}{\pi^{3/2}} \int_{-\infty}^\infty
\frac{\exp(-x^2)}{(\nu-x)^2+a^2} dx
\nonumber \\
 & = & \frac{1}{\sqrt{\pi}} H(a,\nu)
\end{eqnarray}
where $a = \Delta\lambda_L/2\Delta\lambda_D$ and
$\nu = (\lambda-\lambda_0)/\Delta\lambda_D$, where $\Delta\lambda_L$ is
the natural (Lorentzian) line width, and $\Delta\lambda_D$ is the
Doppler shift due to the most probable speed.

We can expect a spectral synthesis code to spend a major part of the
execution time repeatedly calculating the Voigt function. Various
methods are available~\cite{thompson,schreier}. Fast methods usually use
asymptotic methods where possible (since the Voigt profile approaches
the Lorentz and Maxwell profiles in the limit), and other approximate
(eg power series) methods elsewhere. Alternately, the Voigt profile can
be calculated by convolution of the Lorentz and Maxwell Doppler
profiles, using Fourier transformation methods for speed.

But, first, we must find the Lorentzian -- the natural line width and
damping -- and Maxwellian widths -- thermal and other motion.

\subsection{Damping}

The most important contribution to damping is collisions with neutral
hydrogen (almost all of which is in the ground state). Damping due to
electrons is 50 times smaller, or even less, and damping due to helium
about 30 times smaller. Other sources can be neglected.

The line broadening theory developed by Anstee and O'Mara~\cite{anstee}
has been shown to be accurate. Collisions with hydrogen atoms in the
photosphere are fast; the impact approximation can be used. The line
width (HWHM) is
\begin{equation}
\Delta\lambda_{\mathrm{coll}} = N \int_0^\infty v f(v) \sigma (v) dv
\end{equation}
where $\sigma$ is the line broadening cross section, and $f(v)$ is the
Maxwellian velocity distribution.

Data calculated using this theory is
available~\cite{barklem2000}, and code to calculate broadening has been
developed~\cite{barklem1998} (available at
\url{http://www.astro.uu.se/~barklem}) which can be used either for
stand-alone calculation, or can be incorporated into a spectral
synthesis program. This code calculates the linewidth per unit hydrogen
atom density for a given temperature.

The total Lorentzian width is the sum of the collisional damping width,
the natural line width, and the stimulated absorption/emission width.
Only the collisional damping width will be important in the photosphere.

\subsection{Small-scale mass motion}

We can expect small-scale turbulent motion to have a Gaussian
velocity distribution (Kolmogorov turbulence). The effect will be the
same as that of thermal motion, and the two can be combined. The average
photospheric microturbulence is 0.845~km/s. We can simply use this
average value as a uniform microturbulence, or a depth-dependent value
can be used.

\subsection{Intermediate-scale mass motion}

Larger scale mass motion, that cannot be treated as Gaussian turbulence,
will occur. A full treatment would require knowledge of the fluid flow
on length scales of $\approx 10$ -- 100km, and extremely difficult
problem. In general, such motion will act so as to Doppler shift the
opacity; this will produce broadening if it varies with depth.

\section{Integrating the radiative transfer equation}

All of the processes that affect the opacity have been considered. The
radiative transfer equation~(\ref{transfer}) can be integrated.
The radiative transfer equation (\ref{transfer}) is well behaved,
provided we solve it in a reasonable manner. In order to avoid the
growth of errors, we integrate
\begin{equation}
I_\lambda(\tau=0) = \int_0^\infty S_\lambda(\tau_\lambda)
\exp(-\tau_\lambda/\mu)\frac{1}{\mu} d\tau_\lambda.
\end{equation}
In practice, we step through the layers of our model atmosphere, from
the top to the bottom, with the step size for the numerical integration
being determined by the spacing in the model atmosphere. By the time the
bottom of the model atmosphere is reached, the contributions to the
emergent intensity are negligible.

\section{Large-scale mass motion---beyond the plane-parallel
approximation}

Large-scale flows prevent the photosphere from being strictly
plane-parallel. Further effects must be considered as corrections to the
simple plane-parallel model. The simplest method is to assume that such
motions have a Gaussian distribution, and will only act to broaden the
emergent spectrum (as opposed to the microturbulence, which will broaden
the line profile at all depths). This, however, is an unsatisfactory
solution. The main motion at this scale is that due to the granulation.
A granular cell is typically about 1000km across, consisting of a hot,
slowly rising, central region, and a cooler, more rapidly falling outer
region.

Even a simple model of granulation can greatly improve the results of
spectral synthesis~\cite{thesis}. If we divide the granule into three
regions: the rising core, the falling outer region, and a transition
region between the two, we can treat each region as purely
plane-parallel. We can assume that the flow velocity in each region
varies with height $h$ as
\begin{equation}
V = V_0 \exp(-h/V_s),
\end{equation}
where $V_s$ is the velocity scale height. We can expect the
microturbulence, since it is driven by the large scale flow, to have the
same depth dependence. A macroturbulent velocity can also be associated
with each region to model inter-granular variation. A suitable granular
model is shown in table~\ref{granule}. Using this granular model,
calculated spectral lines closely match observed spectral lines in
width and asymmetry.

\begin{table}[htb]
\begin{tabular}{cccc}
Parameter    & Upflow & Transition & Downflow \\
\hline
Area         & 0.45   & 0.40       & 0.15     \\
Brightness   & 1.07   & 0.97       & 0.87     \\
Weighting    & 0.48   & 0.39       & 0.13     \\
$V_0$        & 0.577  & 0          & -1.072   \\
$V_s$        & 368    & 368        & 368      \\
micro        & 1.58   & 3.67       & 3.67     \\
macro        & 1.6    & 1.6        & 3.5      \\
\hline
\end{tabular}
\caption{Granular model. All velocities are in km/s, and heights in km}
\label{granule}
\end{table}

\section{Spectral synthesis}

If we follow the steps outlined here, we can calculate the emergent
spectrum of the sun. The matching of observed and calculated line
profiles allows us to investigate the processes responsible for the
strength and shape of the line---abundance, damping, granulation, etc.
Spectral synthesis, a valuable tool for studying the
atmospheres of the sun and other stars, can be readily implemented on a
typical PC. A number of spectral synthesis codes are
available~\cite{rossSYN,nieminenSYN}.

\end{document}